\documentclass[twocolumn]{aastex63}

\usepackage[fleqn]{amsmath}
\renewcommand{\ng}{NGC\,4051 }
\newcommand{\ngs}{NGC\,4051}
\newcommand{\ngo}{NGC\,4151 }
\newcommand{\ngos}{NGC\,4151}
\newcommand{\h}{$H_0$ }

\newcommand{\hst}{{\it HST} }

\newcommand{\hstw}{{\it F350LP} }
\newcommand{\hstv}{{\it F555W} }
\newcommand{\hsti}{{\it F814W} }
\newcommand{\hsth}{{\it F160W} }
\newcommand{\hstws}{{\it F350LP}}
\newcommand{\hstvs}{{\it F555W}}
\newcommand{\hstis}{{\it F814W}}
\newcommand{\hsths}{{\it F160W}}

\newcommand{\B}{{\it B}-band}                                                             
\newcommand{\V}{{\it V}-band}                                                             
\newcommand{\R}{{\it R}-band}                                                             
\newcommand{\I}{{\it I}-band} 
   
\newcommand{\ignore}[1]{}

\newcommand{\ergsec}{\mbox{erg\,s$^{-1}$}}

\newcommand{\msun}{\mbox{$M_\odot$}}

\newcommand{\hb}{\mbox{\rm H$\beta$}}

\usepackage{xcolor}

\makeatletter
\newcommand*{\rom}[1]{\expandafter\@slowromancap\romannumeral #1@}
\makeatother
\usepackage[normalem]{ulem}

\shorttitle{Distance to the Narrow-Line Seyfert Galaxy \ngs}
\shortauthors{Yuan et al.}

\begin{document}
\title{The Cepheid Distance to the Narrow-Line Seyfert 1 Galaxy NGC\,4051}

\author[0000-0001-9420-6525]{W.~Yuan}
\affiliation{Department of Physics \& Astronomy, Johns Hopkins University, Baltimore, MD 21218, USA}

\author[0000-0002-1775-4859]{L.~M.~Macri}
\affiliation{George P.\ and Cynthia W.\ Mitchell Institute for Fundamental Physics \& Astronomy, Department of Physics \& Astronomy,\\ Texas A\&M University, College Station, TX 77843, USA}

\author[0000-0001-6481-5397]{B.~M.~Peterson}
\affiliation{Department of Astronomy, The Ohio State University, 140 W 18th Ave, Columbus, OH 43210, USA}
\affiliation{Center for Cosmology and AstroParticle Physics, The Ohio State University, 191 West Woodruff Ave, Columbus, OH 43210, USA}

\author[0000-0002-6124-1196]{A.~G.~Riess}
\affiliation{Department of Physics \& Astronomy, Johns Hopkins University, Baltimore, MD 21218, USA}
\affiliation{Space Telescope Science Institute, 3700 San Martin Drive, Baltimore, MD 21218, USA}

\author[0000-0002-9113-7162]{M.~M.~Fausnaugh}
\affiliation{Department of Astronomy, The Ohio State University, 140 W 18th Ave, Columbus, OH 43210, USA}
\affiliation{\ignore{}Kavli Institute for Space and Astrophysics Research,  Massachusetts Institute of Technology,\\ 
77 Massachusetts Avenue, Cambridge, MA 02139, USA}

\author[0000-0002-4312-7015]{S.~L.~Hoffmann}
\affiliation{Space Telescope Science Institute, 3700 San Martin Drive, Baltimore, MD 21218, USA}

\author[0000-0002-5259-2314]{G. S. Anand}
\affiliation{Institute for Astronomy, University of Hawaii, 2680 Woodlawn Drive, Honolulu, HI 96822, USA}

\author[0000-0002-2816-5398]{M.~C.~Bentz}
\affiliation{\ignore{Georgia}Department of Physics and Astronomy, Georgia State University, 25 Park Place, Suite 605, Atlanta, GA 30303, USA}

\author[0000-0001-9931-8681]{E.~Dalla~Bont\`{a}}
\affiliation{\ignore{Padova}Dipartimento di Fisica e Astronomia ``G. Galilei,'' Universit\`{a} di Padova, Vicolo dell'Osservatorio 3, I-35122 Padova, Italy}
\affiliation{\ignore{INAF}INAF-Osservatorio Astronomico di Padova, Vicolo dell'Osservatorio 5 I-35122, Padova, Italy}

\author[0000-0003-4949-7217]{R.~I.~Davies}
\affiliation{\ignore{MPE}Max Planck Institut f\"ur extraterrestrische Physik, Postfach 1312, D-85741, Garching, Germany}

\author[0000-0003-3242-7052]{G.~De~Rosa}
\affiliation{Space Telescope Science Institute, 3700 San Martin Drive, Baltimore, MD 21218, USA}

\author[0000-0002-8224-1128]{L.~Ferrarese}
\affiliation{\ignore{Victoria}NRC Herzberg Astronomy and Astrophysics, National Research Council, 5071 West Saanich Road, Victoria, BC V9E 2E7, Canada}

\author[0000-0001-9920-6057]{C.~J.~Grier}
\affiliation{Department of Astronomy, The Ohio State University, 140 W 18th Ave, Columbus, OH 43210, USA}
\affiliation{\ignore{Steward}Steward Observatory, University of Arizona, 933 North Cherry Avenue, Tucson, AZ 85721, USA}

\author[0000-0002-4457-5733]{E.~K.~S.~Hicks}
\affiliation{\ignore{Anchorage}Department of Physics and Astronomy, University of Alaska Anchorage, AK 99508, USA}

\author[0000-0003-0017-349X]{C.~A.~Onken}
\affiliation{\ignore{ANU}Research School of Astronomy and Astrophysics, Australian National University, Canberra, ACT 2611, Australia}
\affiliation{Australian Research Council (ARC) Centre of Excellence in All-sky Astrophysics (CAASTRO)}

\author[0000-0003-1435-3053]{R.~W.~Pogge}
\affiliation{Department of Astronomy, The Ohio State University, 140 W 18th Ave, Columbus, OH 43210, USA}
\affiliation{Center for Cosmology and AstroParticle Physics, The Ohio State University, 191 West Woodruff Ave, Columbus, OH 43210, USA}

\author[0000-0003-1772-0023]{T.~Storchi-Bergmann}
\affiliation{\ignore{UFRGS}Departamento de Astronomia, Universidade Federal do Rio Grande do Sul, Av. Bento Goncalves 9500, 91501 Porto Alegre, RS, Brazil}

\author[0000-0001-9191-9837]{M.~Vestergaard}
\affiliation{\ignore{Dark}DARK Niels Bohr Institute, University of Copenhagen, Jagtvej 128, 2200 Copenhagen N, Denmark}
\affiliation{\ignore{Steward}Steward Observatory, University of Arizona, 933 North Cherry Avenue, Tucson, AZ 85721, USA}

\begin{abstract}

We derive a distance of $D = 16.6 \pm 0.3$~Mpc ($\mu=31.10\pm0.04$~mag) to the archetypal narrow-line Seyfert 1 galaxy \ng based on Cepheid Period--Luminosity relations and new {\it Hubble Space Telescope} multiband imaging. We identify 419 Cepheid candidates and estimate the distance at both optical and near-infrared wavelengths using subsamples of precisely-photometered variables (123 and 47 in the optical and near-infrared subsamples, respectively). We compare our independent photometric procedures and distance-estimation methods to those used by the SH0ES team and find agreement to 0.01~mag. The distance we obtain suggests an Eddington ratio $\dot{m} \approx 0.2$ for \ngs, typical of narrow-line Seyfert 1 galaxies, unlike the seemingly-odd value implied by previous distance estimates. We derive a peculiar velocity of $-490\pm34$~km~s$^{-1}$ for \ngs, consistent with the overall motion of the Ursa Major Cluster in which it resides. We also revisit the energetics of the \ng nucleus, including its outflow and mass accretion rates.\\
\end{abstract}

\section{Introduction}
The distance to an astronomical object is one of the most fundamental measurements that can be made, and yet it is also one of the most difficult. Without an accurate distance, measurements of many other parameters are of limited utility. In the case of the nearest active galactic nuclei (AGNs),  the lack of accurate and precise distance measurements to these objects precludes a robust calibration of their physical properties, even though, compared to the general AGN population, their detailed structure is best resolved and their dynamics and energetics are the most easily studied.  Thus, obtaining accurate distances is the most urgent priority toward achieving an accurate physical description of the accretion, feedback, and energetics of nearby AGNs.

\ngs, one of the original six Seyfert galaxies \citep{Seyfert1943} and one of the lowest-redshift AGNs ($z = 0.00234$), illustrates some of the difficulties. \ng is the archetype of the subclass of AGNs known as ``narrow-line Seyfert 1 galaxies'' \citep[NLS1s;][]{Osterbrock85}, which are thought to be high Eddington ratio objects; the Eddington ratio $\dot{m}$ is defined as the ratio of mass accretion rate to the Eddington accretion rate, $\dot{m} = \dot{M}/\dot{M}_\mathrm{Edd}$. Since the accretion rate is $\dot{M} = L_\mathrm{bol}/\eta c^2$ where $L_\mathrm{bol}$ is the bolometric luminosity and $\eta$ is the efficiency, as long as $\eta$ is independent of $\dot{M}$, we can also write $\dot{m} = L_{\rm bol}/L_\mathrm{Edd}$ where $L_\mathrm{Edd}$ is the Eddington luminosity. To compute the bolometric luminosity, we must know the distance. Unfortunately, redshift-independent distances for \ng are ambiguous. Distances quoted in the literature cover a wide range, from a low of 8.8\,Mpc \citep{Sorce2014} to a high close to 18\,Mpc \citep{Wang2014,2014ApJ...784L..11Y}. If \ng is at the lower end of this range, which is close to the redshift-based distance of 9.5\,Mpc, its Eddington ratio is surprisingly low for an NLS1 galaxy. By using the bolometric correction and weighted mean mass and luminosity from \citet{DB2020}, we find that at a distance of 9\,Mpc the Eddington ratio would be $\dot{m} \approx 0.07$, which is fairly typical of other local Seyfert galaxies. However, if \ng is at the high end of this distance range, at 18\,Mpc, $\dot{m} \approx 0.28$, which is more typical of NLS1s.

\ng is the first AGN for which strong evidence for inflows in molecular gas was found using near-infrared integral field spectroscopy \citep{Riffel2008}.  The inferred geometry of the inflow depends strongly on the distance to the galaxy, taken to be 9.3\,Mpc in that work. Similarly, X-ray spectra reveal a fast ionized outflow that loses kinetic energy by interaction with the interstellar medium in the host galaxy \citep{Pounds2014}; the energetics and timescales again strongly depend on the distance to the AGN and are consequently systematically uncertain by a factor of several. Moreover, spatially resolved observations of the narrow-line region reveal a strong outflow component \citep{Fischer2013}, and again the energetics are uncertain on account of the poorly constrained distance. The quantitative results of these observations are important because they are probes of the feedback mechanisms through which supermassive black hole (SMBH) accretion can modify the interstellar medium and regulate star formation in the host galaxy. Understanding the physical mechanisms at work in both feeding and feedback, and determining whether or not the feedback mechanisms of theoretical models actually work in nature, requires detailed observations of the complex inner structure of AGNs. Nearby galaxies afford the only opportunity to test in detail --- i.e., on sub-hundred pc scales --- the prescriptions used in models of galaxy and SMBH co-evolution. It is on scales of less than 100 pc that morphological differences between active and quiescent galaxies appear \citep{Lopes2007,GB2012,Hicks2013,Davies2014}. There is observational and theoretical evidence that disk processes are important in feeding AGNs across broad ranges of redshift and luminosity \citep[e.g.,][]{Hopkins2010, Schawinski2012,Kocevski2012,Riffel2013,SM2014,Storchi2019}, so what we learn about local AGNs is applicable to galaxies at $z > 1$, where co-evolution largely occurs.

Finally, it is only in the nearest AGNs that we can compare SMBH masses measured from stellar or gas dynamical modeling, which are distance-dependent, with those measured by reverberation mapping (RM), which are distance-independent. \ng is challenging as the SMBH radius of influence ($r_{\rm BH} = GM/\sigma_*^2$, where $M$ is the BH mass and $\sigma_*$ is the stellar bulge velocity dispersion) is $\sim 0.67$\,pc; whether or not it is spatially resolvable with either the {\it Hubble} or the {\it James Webb\ } Space Telescopes (HST, JWST) is unclear given the factor of 2--3 uncertainty in the SMBH mass and the factor of two uncertainty in the distance, but it should nevertheless be possible to get at least meaningful constraints on the central mass with the IFU on {\it JWST}. 

Our goal is to eliminate distance-dependent ambiguities by determining a Cepheid-based distance to \ngs. The Cepheid Period--Luminosity Relation (PLR) or Leavitt Law \citep{Leavitt1912} is a widely-used empirical standard candle with a reach of at least $D \approx 40$~Mpc \citep{Riess2016}. It has been extensively applied to dozens of extragalactic systems, including \citet{Bentz2019} and \citet[][hereafter Y20]{Yuan2020} who recently measured the Cepheid-based distances to the Seyfert 1 galaxies NGC\,6814 and \ngos, respectively.

In this work, we derive a Cepheid distance to \ng using multiband \hst observations. The rest of this paper is organized as follows. We present the observations, data reduction, and photometry in \S2 and our Cepheid selection methods in \S3. Our main result, the Cepheid distance to \ngs, is given in \S4 and we discuss its implications in \S5.

\section{Observations, Data Reduction, and Photometry}

\begin{figure*}
\includegraphics[width=\textwidth]{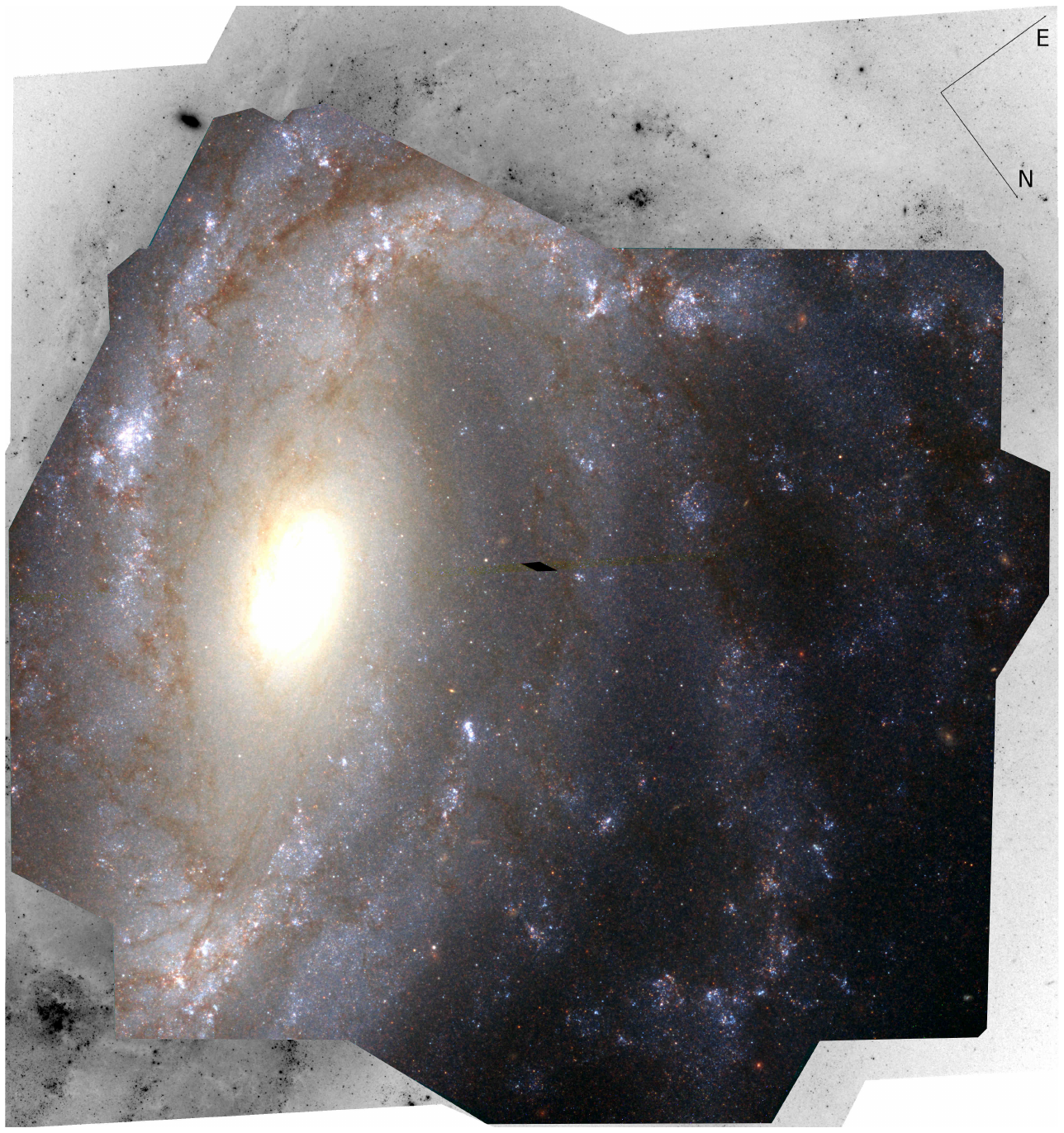}
\caption{Pseudo-color image of \ng using our master frames in \hstws, \hstvs, \hstis, and \hsth as the luminance layer and the blue, green, and red channels, respectively. A square-root scaling was applied to all layers to increase the visibility. Color is only shown where all layers overlap, while an inverse grayscale is used in areas not imaged with \hsths. The compass lines are $20\arcsec$ in length. The black diamond-shaped area near the center is due to the gap between the WFC3/UVIS CCDs.\label{fig_cc}}
\end{figure*}

We observed \ng and carried out the subsequent analysis following the same strategy used by Y20 for \ngos. We obtained optical and near-infrared (NIR) \hst images of a field that covers a large area of the disk (see Figure~\ref{fig_cc}). We briefly summarize the observations, data reduction, and photometry procedure below and refer interested readers to Y20 for further details.

The \ng observations include 12, 3, 3, and 6 epochs in \hstws, \hstvs, \hstis, and \hsths, respectively. The observation baseline spans 73 days, during which the orientation angle of \hst changed once. The observation log and exposure times are given in Table~\ref{tab_obs}. We note that one of the two \hstv images obtained in epoch 9 had a significantly worse FWHM, possibly due to increased jitter, and thus was excluded from further analysis.

\begin{deluxetable}{cccccc}
\tablecaption{Observation Log\label{tab_obs}}
\tablewidth{0pt}
\tablehead{
\colhead{Epoch} & \colhead{MJD} & \multicolumn{4}{c}{Dither$\times$Exposure Time (seconds)} \\ \cline{3-6}
&&\hstw & \hstv & \hsti & \hsth
}
\startdata
  1 &   58089.2 & 3$\times$350 & \nodata & \nodata & 2$\times$553 \\
  2 &   58107.8 & 3$\times$350 & 2$\times$550 & \nodata & \nodata \\
  3 &   58111.9 & 3$\times$350 & \nodata & 2$\times$550 & \nodata \\
  4 &   58117.4 & 3$\times$350 & \nodata & \nodata & 2$\times$553 \\
  5 &   58122.6 & 3$\times$350 & 2$\times$550 & \nodata & \nodata \\
  6 &   58129.0 & 3$\times$350 & \nodata & \nodata & 2$\times$553 \\
  7 &   58135.6 & 3$\times$350 & \nodata & 2$\times$550 & \nodata \\
  8 &   58141.0 & 3$\times$350 & \nodata & \nodata & 2$\times$553 \\
  9 &   58145.3 & 3$\times$350 & 2$\times$550$^a$ & \nodata & \nodata \\
 10 &   58149.6 & 3$\times$350 & \nodata & \nodata & 2$\times$553 \\
 11 &   58155.2 & 3$\times$350 & \nodata & 2$\times$550 & \nodata \\
 12 &   58162.1 & 3$\times$350 & \nodata & \nodata & 2$\times$553
\enddata
\tablecomments{$a$: The first image of this set was excluded due to its poor FWHM, possibly due to increased jitter.}
\end{deluxetable}

\begin{deluxetable}{lllll}
\tabletypesize{\scriptsize}
\tablecaption{Secondary Standards\label{tab_sec}}
\tablewidth{0pt}
\tablehead{
\colhead{R.A.} & \colhead{Dec.} & \colhead{\hstw} & \colhead{\hstv} & \colhead{\hsti} \\ \cline{3-5}
\multicolumn{2}{c}{[J2000]} & \multicolumn{3}{c}{[mag(mmag)]}
}
\startdata
180.76803 &    44.53942 &   23.520(13) &   23.604(18) &   22.862(16) \\
180.77227 &    44.53125 &   23.943(15) &   24.202(21) &   23.034(14) \\
180.77507 &    44.52808 &   23.601(23) &   23.707(19) &   23.372(30) \\
180.77511 &    44.53187 &   23.566(17) &   23.772(20) &   22.798(20) \\
180.77630 &    44.53949 &   23.089(21) &   23.228(20) &   23.009(30) \\
180.77693 &    44.54960 &   23.822(19) &   24.041(23) &   22.981(21) \\
180.78067 &    44.54180 &   23.060(18) &   23.309(23) &   22.898(24) \\
180.78152 &    44.54246 &   22.434(11) &   22.603(13) &   21.713(11) \\
180.78185 &    44.51941 &   22.824(20) &   22.894(15) &   22.462(28) \\
180.78258 &    44.54943 &   23.771(18) &   23.959(24) &   23.325(31)
\enddata
\tablecomments{Coordinates are based on the WCS solution of the first \hstw image. Uncertainties are given in parentheses and are expressed in units of mmag. This table is available in its entirety in machine-readable form.}
\end{deluxetable}

We registered and drizzle-combined all images using the {\tt AstroDrizzle v2.2.6} package. We created master frames in each band by combining the respective images obtained in all epochs. We also created drizzled images of each \hstw epoch for variable detection. We modeled the surface brightness gradients across all images and subtracted them using the same methods as described by Y20.

We carried out point-spread function (PSF) photometry on the \ng images following identical procedures to those used in the case of \ngo (Y20). We firstly derived a source list through a two-pass source detection on the \hstw master frame using {\tt DAOPHOT/ALLSTAR} \citep{Stetson1987}, then performed {\tt ALLFRAME} \citep{Stetson1994} time-series photometry on the 12 epochs of images obtained through that filter. The frame-to-frame magnitude offsets were calculated based on 69 carefully-selected secondary standards which are listed in Table~\ref{tab_sec}. We transformed the coordinates of the objects in the \hstw source list to carry out fixed-position PSF photometry in the \hstvs, \hsti \& \hsth master frames (no time-series measurements were obtained, only mean magnitudes). We performed an additional detection step on the \hsth master frame after removing the stars that were present in the \hstw source list.

\section{Cepheid Identification and Characterization}

We applied the same Cepheid selection procedures as described by Y20 and identified 419 Cepheid candidates in \ngs. We measured their periods, amplitudes, and crowding bias corrections.

\subsection{Cepheid identification}\label{sec_cep}

\begin{figure}[!t]
\includegraphics[width=0.49\textwidth]{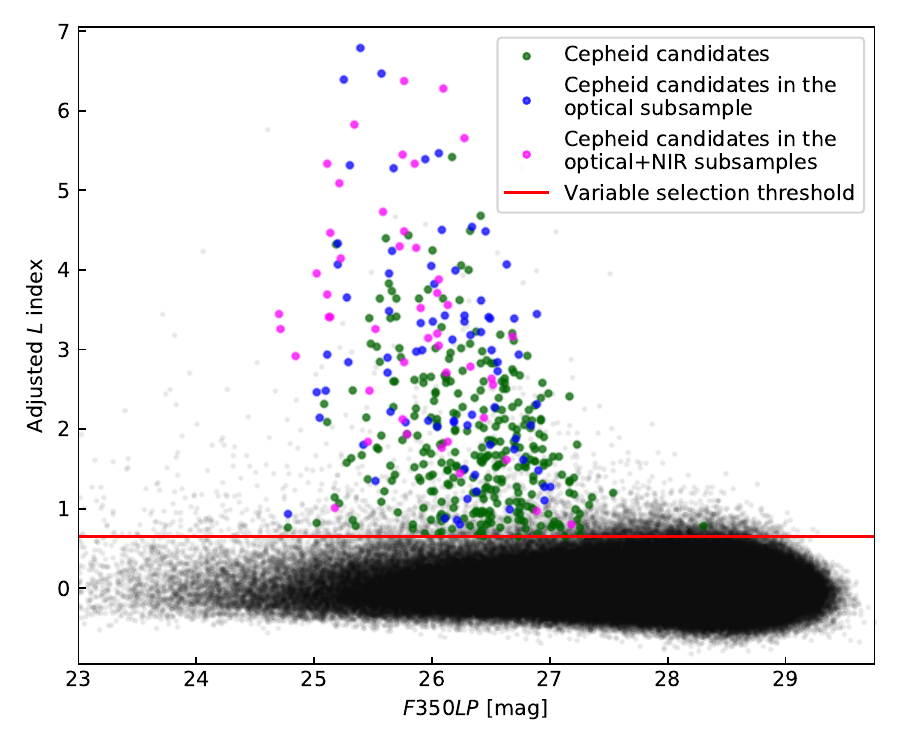}
\caption{Adjusted $L$ index against \hstw magnitudes for all the sources. The red line indicates our threshold for variable selection. Larger points indicate the 419 Cepheid candidates described in \S\ref{sec_cep}. The NIR subsample described in \S\ref{sec_sub} is indicated by magenta points, while the optical subsample described in \S\ref{sec_sub} is indicated by both magenta and blue points.\label{fig_lm}}
\ \par
\end{figure}

\begin{figure}[!t]
\includegraphics[width=0.49\textwidth]{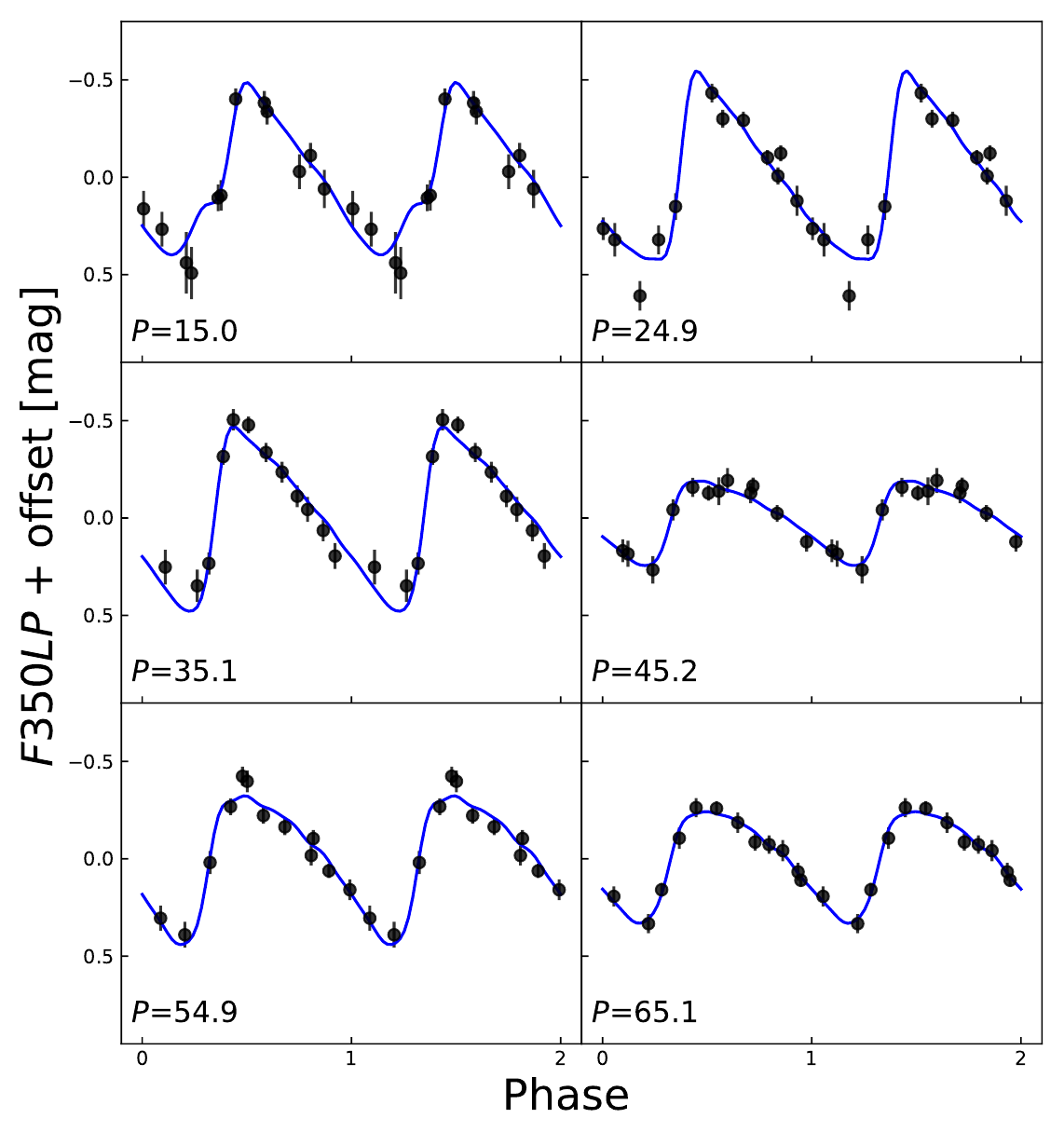}
\caption{Example \hstw light curves of selected Cepheids spanning the period range. The blue curves indicate the best-fit Cepheid templates. We plot two cycles of pulsation for visualization purposes.\label{fig_lc}}
\end{figure}

Our Cepheid identification included four steps: variability index cut \citep[][adjusted for magnitude dependence]{Stetson1996}, template fitting, period and amplitude cuts, and visual inspection. We firstly selected 4519 variable candidates with Stetson index $L>0.65$. This threshold was slightly higher than the initial choice for \ngo ($L>0.5$) for two reasons: (1) the minimum $L$ value for \ngo Cepheids that were eventually used for distance determination was $\sim$0.9; (2) the \ng field contains more Cepheid-yielding star-forming regions, and thus the number of variables is not a limiting factor for the \ng distance determination. Our goal is not to discover the complete set of Cepheids in \ng but rather to identify a high-purity sample that enables a precise and accurate distance determination. The $L$--magnitude distribution for \ng and our updated $L$ threshold are shown in Figure~\ref{fig_lm}. Secondly, we fit the \citet{Yoachim2009} Cepheid templates to the \hstw light curves of these variable candidates using the method described by Y20. Thirdly, we restricted the sample based on the best-fit periods and amplitudes as done in Y20; this yielded 2804 objects with $P<90$d and peak-to-trough amplitudes between 0.4 and 1.7~mag. Finally, we visually inspected the phase-folded \hstw light curves of these 2804 objects and their overall matches to the best-fit Cepheid templates by overplotting measurements and templates. We identified 419 Cepheid candidates through this visual inspection process. Rejected objects are mostly non-variables with noisy measurements which passed our initial selection criteria. Light curves of all 2804 pre-selected objects are available as supplementary material. Figure~\ref{fig_lc} shows 6 examples to illustrate the light curve quality of the selected variables. The \hstw light curves of all 419 visually selected objects are given in Table~\ref{tab_lc}. Our motivation for this inspection is to obtain a high-purity sample, which is superior to a larger but possibly contaminated sample for distance measurement. Though subjective, visual inspection remains an effective and widely-adopted method for the final Cepheid confirmation in extragalactic distance studies \citep[e.g.,][]{1996ApJ...466...55S,1998ApJ...508..491S,2015MNRAS.450.3597F,2015AJ....149..183H,Hoffmann2016,Bentz2019}.

\begin{deluxetable}{lrcc}
\tablecaption{\hstw light curves\label{tab_lc}}
\tablewidth{0.49\textwidth}
\tablehead{
\colhead{Cepheid} & \colhead{t$^a$} & \colhead{mag} & \colhead{$\sigma$} \\ \cline{3-4}
\colhead{ID} & [day] & \multicolumn{2}{c}{[mag]}
}
\startdata
     9734 &   0.729 &    26.518 &    0.062 \\
     9734 &  19.282 &    26.473 &    0.070 \\
     9734 &  23.394 &    26.610 &    0.055 \\
     9734 &  28.883 &    26.705 &    0.079 \\
     9734 &  34.113 &    26.199 &    0.045 \\
     9734 &  40.468 &    26.267 &    0.067 \\
     9734 &  47.088 &    26.703 &    0.111 \\
     9734 &  52.458 &    26.353 &    0.053 \\
     9734 &  56.823 &    26.135 &    0.051 \\
     9734 &  61.067 &    26.201 &    0.051 \\
     9734 &  66.699 &    26.537 &    0.051 \\
     9734 &  73.602 &    26.595 &    0.078
\enddata
\tablecomments{$a$: JD$-$2458089. This table is available in its entirety in machine-readable form.}
\end{deluxetable}

Thanks to the template-fitting technique, Cepheid periods can be effectively recovered with sparsely sampled light curves. Typically, $\sim$12-epoch optical observations are designed to search for Cepheids in extragalactic systems and measure their periods \citep{1991PASP..103..933M}. Because there are only 4 free parameters (including the period) in the model, it is relatively easy to recover the Cepheid periods with $\sim$12 measurements. However, we note that the light curve sampling requires a span of $\sim$60-90 days and nonredundant spacings to ensure adequate phase coverage \citep{2005ApJ...630.1054M}. In this work, we solved the Cepheid periods using a nonlinear least-square fitting method, with the exact methodology described in Y20. To ensure the global least-square fit is reached, we included a grid of 150 logarithmically-spaced initial trial periods from 10 to 120 days. We note that the fitting algorithm is allowed to converge to any period within the entire range, regardless of the starting trial value. As a result, for a given object, the 150 initial trial periods usually converged to only a few final values where  $\chi^2$ reached local minima. We adopted the output period with the lowest $\chi^2$ as the best fit. For the 419 Cepheid candidates, we further estimated their period accuracy using the bootstrapping method. We computed the template-fitting residuals for each light curve, resampled and added them back to the original light curve, and then determined the residual-resampled light curve period. This process was repeated 200 times, and the standard deviation of the resulting periods was adopted as the period uncertainty. We found a mean (median) relative period uncertainty $\Delta P/P$ of 2.0\% (1.7\%). The occurrences of $|\Delta P/P| > $ 0.05 and 0.1 were 4\% and 0.5\%, respectively. Their contribution to the final distance error budget is studied in \S\ref{sec_eb}.

We considered whether the measurement uncertainty in period combined with the greater frequency of short period than long period Cepheids could bias the period determination.  Using a realistic distribution of periods we found for the mean case where $\sigma_{\log P} \sim 0.007$,  the bias in the inferred period is  about 2\% of the period measurement uncertainty or $\Delta\log P$=0.00015 and combined with the PL relation a bias in the inferred magnitude of $\sim$ 0.0004 mag.  For Cepheids with periods closer to the span of the observations, $\log P \sim$ 1.8, the mean bias rises to 5\% of the uncertainty or $\Delta \log P$=0.00035, still quite negligible.  However, this bias can become important when the measurement uncertainty in $\log P$ exceeds 0.05 and the bias in the PLR approaches 0.01 mag.

\subsection{Crowding correction and magnitude calibration}\label{sec_mag}

We applied crowding, aperture, and phase corrections to the Cepheid magnitudes, then calibrated our photometric measurements on the Vega system.

While we adopted the same crowding correction method as for \ngos, we slightly updated our procedure in \hsth relative to Y20. In our previous work, we only measured the \hsth magnitude and crowding bias for Cepheids with $P>25$~d. We removed such a restriction for this analysis; 379 of the optically discovered Cepheids were located in the smaller NIR field of view. Of these, 278 were detected and photometered. We determined their \hsth crowding correction $\Delta\mathrm{\hsth}$ by carrying out three passes of the artificial star test procedure, compared to only one pass for \ngos. In the first pass, we determined the artificial star magnitudes using a PLR (with a slope fixed to $-3.2$ mag/dex) based on Cepheids with $P>40$~d. In the next two passes, we based our calculation on Cepheids with $\sigma(\Delta\mathrm{\hsth}) < 0.25$~mag as determined from the preceding pass. The restriction of $P>40$~d and $\sigma(\Delta\mathrm{\hsth}) < 0.25$~mag yielded a more accurate estimation of the input artificial star magnitudes, and our tests converged faster.

\begin{figure}
\includegraphics[width=0.49\textwidth]{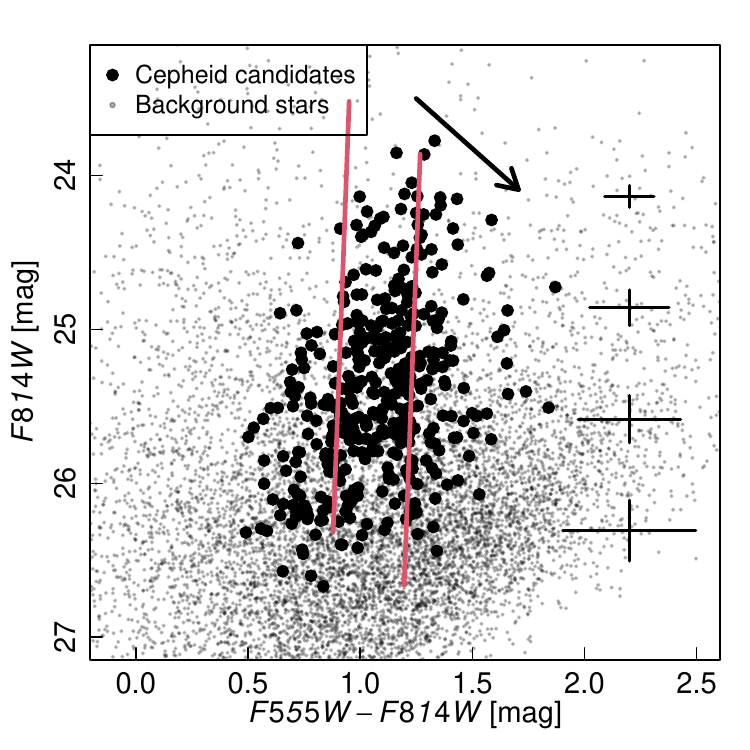}
\caption{Color--magnitude diagram of optically-selected Cepheids (filled symbols). Representative errorbars are shown on the right. Gray dots are a subsample of field stars. The area between the red lines shows the Cepheid instability strip for $P\!>\!10$~d, derived from LMC observations \citep{Riess2019}. The arrow indicates the effect of $A_V=1$~mag.\label{fig_cmd}}
\end{figure}

\begin{figure}
\includegraphics[width=0.49\textwidth]{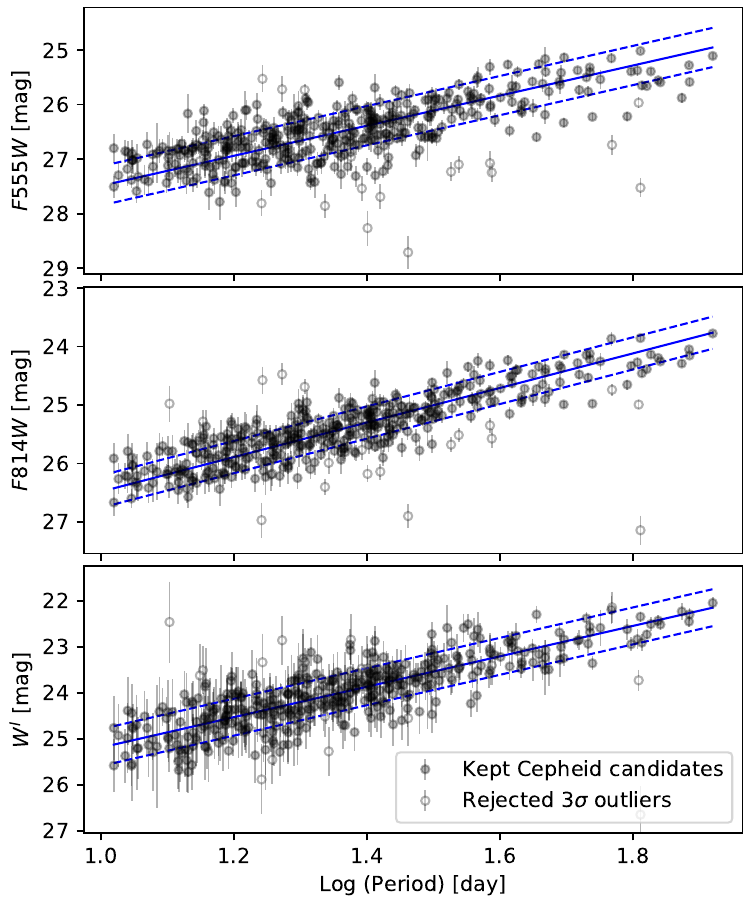}
\caption{\hstv (top), \hsti (middle), and $W^I$ (bottom) PLRs for optically selected Cepheids (filled circles). Rejected variables are indicated by open circles. The solid and dashed blue lines indicate the best-fit PLRs and $\pm1\sigma$ scatter, respectively.\label{fig_oplr}}
\end{figure}

\begin{deluxetable*}{ccccccccccccccc}[!t]
\tabletypesize{\tiny}
\tablecaption{Properties of Cepheid candidates \label{tab_ceph}}
\tablewidth{0pt}
\tablehead{
\colhead{ID} & \colhead{R.A.} & \colhead{Dec.} & \colhead{$P^b$} & $L$ & \colhead{Amp.} & \colhead{\hstw} & \colhead{\hstv} & \colhead{\hsti} & \colhead{\hsth} & $SB^d$ & F$^e$ & \colhead{$\Delta$\hstvs} & \colhead{$\Delta$\hsti} & \colhead{$\Delta$\hsth} \\ \cline{2-3} \cline{7-10}  \cline{13-15}
&\multicolumn{2}{c}{[J2000]$^a$}&[day]&&[mag]&\multicolumn{4}{c}{[mag (mmag)]$^c$} & & & \multicolumn{3}{c}{[mmag]$^f$}
}
\startdata
   82977 &   180.769100 &    44.538239 &      24.4(0.4) &      1.69 &     0.938 &    26.236(103) &    26.742(132) &    25.362(150) &    24.062(601) &    308.0 &   &   -7(106) &    3(120) &  518(587) \\
   88413 &   180.769205 &    44.537613 &      13.4(0.1) &      2.27 &     1.157 &    26.617(133) &    26.843(166) &    26.142(239) &    26.284(573) &    295.3 &   &   57(145) &   59(205) & 1308(508) \\
  142230 &   180.769992 &    44.531406 &      14.8(0.1) &      3.63 &     0.885 &    26.257(158) &    26.506(181) &    25.572(184) &        \nodata &    253.9 &   &   33(148) &   -3(165) &   \nodata \\
  129685 &   180.771219 &    44.533581 &      14.0(0.2) &      0.86 &     0.479 &    26.440(113) &    26.759(145) &    25.610(178) &    25.161(638) &    251.5 &   &   13(127) &   28(152) &  926(614) \\
  125827 &   180.771822 &    44.534374 &      76.8(7.1) &      3.70 &     0.436 &    25.147( 36) &    25.584( 69) &    24.152( 46) &    22.963(222) &    275.9 & H &    7( 37) &    9( 36) &    5(212) \\
  115396 &   180.771970 &    44.535775 &      20.2(0.4) &      0.71 &     0.830 &    26.049(171) &    26.450(202) &    25.146(292) &    23.554(601) &    304.2 &   &   67(156) &   20(209) &  692(592) \\
  101787 &   180.772327 &    44.537594 &      58.6(2.2) &      0.95 &     0.418 &    24.757(105) &    25.149( 93) &    23.863(120) &    22.090(288) &    329.8 & I &  -13( 66) &   10( 69) &   48(275) \\
  126388 &   180.772356 &    44.534583 &      11.0(0.1) &      0.69 &     1.306 &    26.866(261) &    27.180(311) &    26.225(280) &        \nodata &    284.8 &   &   11(272) &   57(240) &   \nodata \\
   42535 &   180.772570 &    44.545120 &      17.8(0.3) &      1.60 &     1.124 &    26.847(129) &    27.190(192) &    25.969(188) &    24.942(602) &    283.4 &   &   38(140) &   20(145) &  836(578) \\
 3059861 &   180.772803 &    44.541936 &      31.3(0.6) &      1.92 &     0.873 &    25.746(225) &    26.104(289) &    25.114(181) &        \nodata &    453.3 &   &  203(280) &  133(153) &   \nodata \\
  116297 &   180.772926 &    44.536114 &      19.9(0.5) &      1.02 &     0.890 &    25.881(125) &    26.134(148) &    25.065(161) &    24.084(505) &    309.5 &   &   25(130) &   -3(128) &  704(484) \\
  113897 &   180.772936 &    44.536426 &      16.0(0.3) &      0.81 &     0.776 &    26.308(171) &    26.774(217) &    25.576(216) &    25.216(530) &    316.8 &   &   14(152) &   10(173) & 1172(504) \\
  131049 &   180.772952 &    44.534324 &      46.7(1.0) &      2.95 &     0.711 &    25.158( 71) &    25.398( 92) &    24.384( 80) &    24.125(427) &    287.4 & I &   29( 79) &   23( 67) &  144(380) \\
  153160 &   180.772990 &    44.531602 &      64.4(4.3) &      3.01 &     0.728 &    25.693( 61) &    25.965( 98) &    24.992( 83) &    23.977(258) &    263.3 & o &    8( 60) &    3( 55) &   -1(234) \\
   38213 &   180.773388 &    44.546094 &      32.8(0.7) &      3.74 &     0.812 &    25.618(118) &    26.034(130) &    24.824(121) &    23.711(491) &    280.3 &   &   20(121) &   38(108) &  315(485) \\
  182138 &   180.773570 &    44.528351 &      26.3(0.7) &      2.29 &     0.773 &    25.642(110) &    25.953(142) &    24.943(135) &    24.022(457) &    259.5 &   &   74(128) &   42(116) &  153(447) \\
   49235 &   180.773713 &    44.544866 &      12.8(0.2) &      0.87 &     0.711 &    26.690(211) &    26.850(246) &    25.791(277) &        \nodata &    293.0 &   &   80(234) &   46(254) &   \nodata \\
   71671 &   180.774197 &    44.542297 &      49.0(0.9) &      1.59 &     0.635 &    25.362(144) &    25.635(173) &    24.609(133) &        \nodata &    430.9 &   &  114(159) &   73( 99) &   \nodata \\
   81958 &   180.774332 &    44.541084 &      26.4(0.1) &      1.48 &     0.994 &    26.225(185) &    26.548(184) &    25.664(174) &        \nodata &    400.3 &   &   -6(147) &   11(145) &   \nodata \\
  201732 &   180.774472 &    44.526389 &      38.5(0.3) &      1.95 &     0.888 &    25.739(104) &    26.046(141) &    24.944(113) &    24.225(342) &    239.1 & H &  -20( 94) &  -20( 91) &   44(276)
\enddata
\tablecomments{$a$: Coordinates from first \hstw image.
$b$: Period uncertainties are shown in the parentheses.
$c$: Fully-calibrated Vega-system magnitudes, including crowding corrections and their uncertainties.
$d$: \hsths-band local surface brightness in units of counts/s/sq arcsec.
$e$: Cepheids used for $W^I$ PLR (I), for both $W^I$ and $W^H$ PLRs (H), or rejected from optical PLRs (o).
$f$: Crowding corrections and uncertainties.
This table is available in its entirety in machine-readable form.}
\end{deluxetable*}

Unlike Y20, we derived aperture corrections using field stars (listed in Table~\ref{tab_sec}) for \hstws, \hstvs, and \hstis. In the case of \hsths, we adopted the aperture correction value derived by Y20 due to the insufficient number of suitable stars in the respective master frame. All magnitudes presented in this paper are in the Vega system and were calibrated using the current STScI zeropoints (see \S4.2 of Y20 for the detailed method). The adopted Vega zeropoints for ``infinite aperture'' magnitudes are 26.817, 25.843, 24.712, and 24.689 for \hstws, \hstvs, \hstis, and \hsths, respectively.

We present the Cepheid periods, amplitudes, fully calibrated magnitudes, and crowding corrections in Table~\ref{tab_ceph}. We adopted the \citet{Riess2019} PLR slopes of -2.76, -2.96, and -3.31 in \hstvs, \hstis, and $W^I$ (defined below), respectively. We iteratively fit these PLRs to our sample, rejecting the single largest $>$3$\sigma$ outlier in each step until convergence. We arrived at a final optically-selected sample of 397 variables. Their color--magnitude diagram and optical PLRs are shown in Figure~\ref{fig_cmd} and Figure~\ref{fig_oplr}, respectively.

\section{Results}

We determined a distance to \ng of $D = 16.62 \pm 0.32$~Mpc, equivalent to a distance modulus of $\mu = 31.103 \pm 0.042$~mag, by applying the reddening-free Wesenheit indices \citep{Madore1982, Riess2019}

\begin{eqnarray}
W^I & = &\mathrm{\hsti} - 1.3 (\mathrm{\hstv} - \mathrm{\hsti}),\\
W^H & = & \mathrm{\hsth} - 0.386 (\mathrm{\hstv} - \mathrm{\hsti})
\end{eqnarray}

\noindent to Cepheid subsamples with precise measurements and averaging the $W^I$ and $W^H$ PLR offsets relative to the LMC. In this section, we present the details of this selection, the corresponding distance determination, and a comparison to results obtained with an alternative procedure \citep{Riess2016}.

\vfill\pagebreak

\subsection{Selection of subsamples for distance determination}\label{sec_sub}

\begin{figure}
\includegraphics[width=0.49\textwidth]{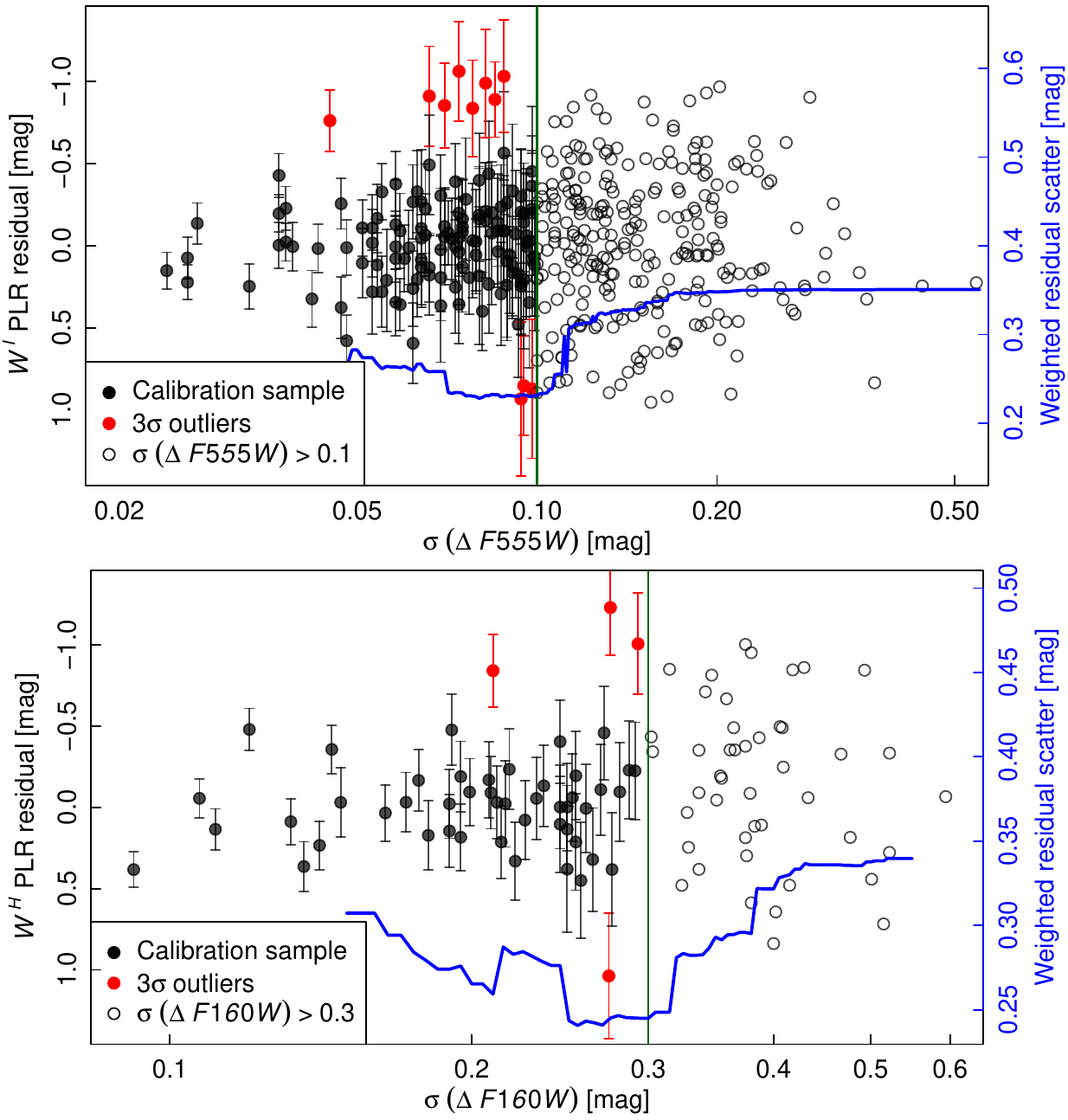}
\caption{Optical (upper) and NIR (lower) subsamples used for distance determination, selected through limits on crowding correction uncertainty and iterative 3$\sigma$ outlier rejection. The selected Cepheids, rejected outliers, and excluded Cepheids are indicated by black filled circles, red filled circles, and gray open circles, respectively. The blue curves show the weighted scatter of PLR residuals as a function of maximum crowding correction uncertainty, while the green vertical lines indicate our adopted cuts.\label{fig_cs}}
\end{figure}

We selected high-quality subsamples of Cepheids for distance determination purposes, in order to obtain tight Cepheid PLRs and reduce the noise from poorly-measured outliers that were not detected in \S\ref{sec_mag}. The leading source of scatter in the $W^I$ index is the uncertainty in the \hstv crowding correction $\sigma(\Delta\mathrm{\hstv})$. Thus, we empirically determined a cut of $\sigma(\Delta\mathrm{\hstv}) < 0.1$~mag based on the weighted scatter of the $W^I$ PLR residuals, as shown in the upper panel of Figure~\ref{fig_cs}. We applied an iterative 3$\sigma$ outlier rejection to obtain a subsample of 123 Cepheids for the optical distance determination. Similarly, we then applied a further empirical cut of $\sigma(\Delta\mathrm{\hsth}) < 0.3$~mag to arrive at a ``NIR distance subsample'' of 47 Cepheids, as shown in the lower panel of Figure~\ref{fig_cs}. The weighted scatter of the full $W^I$ and $W^H$ samples are 0.35 and 0.41~mag, respectively, while the corresponding values for the distance-determination subsamples are 0.23 and 0.24 mag, respectively.

\begin{figure}
\begin{interactive}{js}{fig07.tar.gz}
\includegraphics[width=0.49\textwidth]{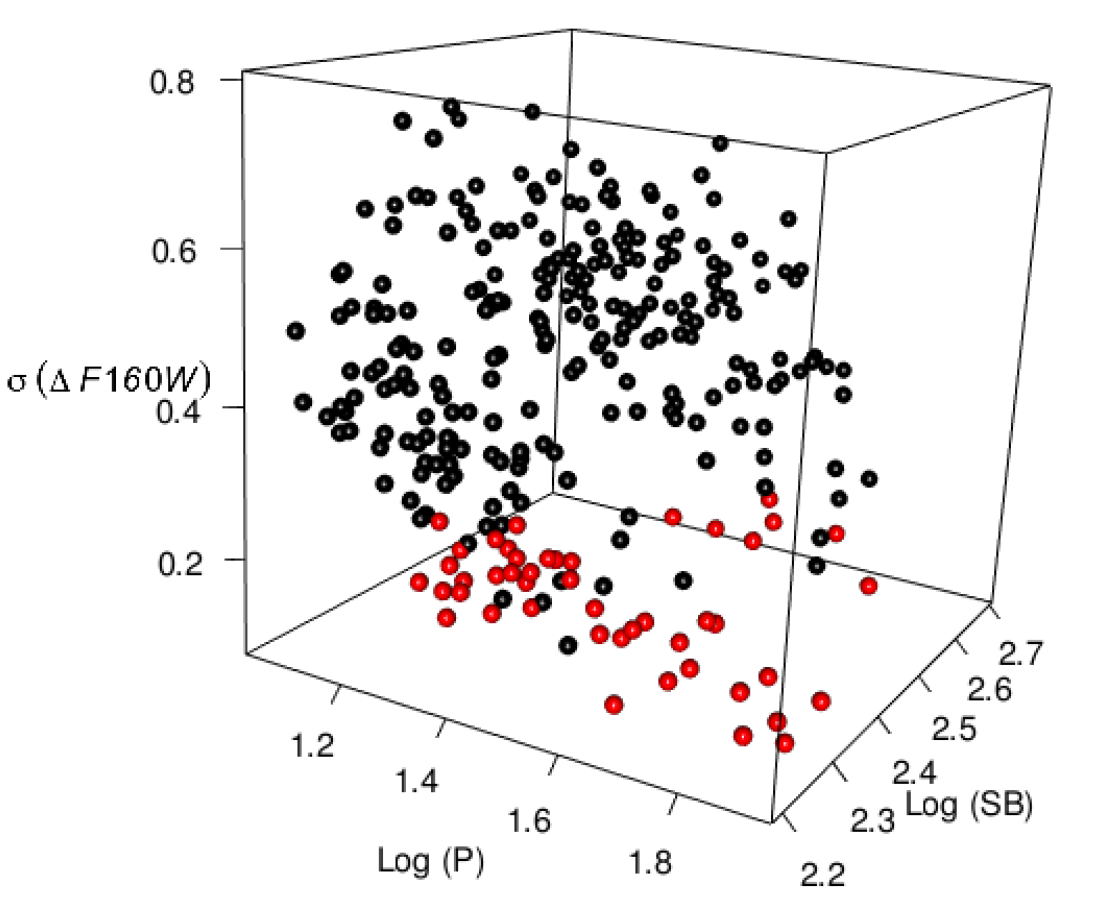}
\end{interactive}
\caption{3-D plot showing the correlation between the pairs among $\sigma(\Delta\mathrm{\hsth})$, period, and local surface brightness (SB). The  NIR subsample used for distance determination is indicated by red points. An interactive version of this plot that allows rotation and zoom is available in the online article.\label{fig_pss}}
\end{figure}

\begin{figure*}
\includegraphics[width=\textwidth]{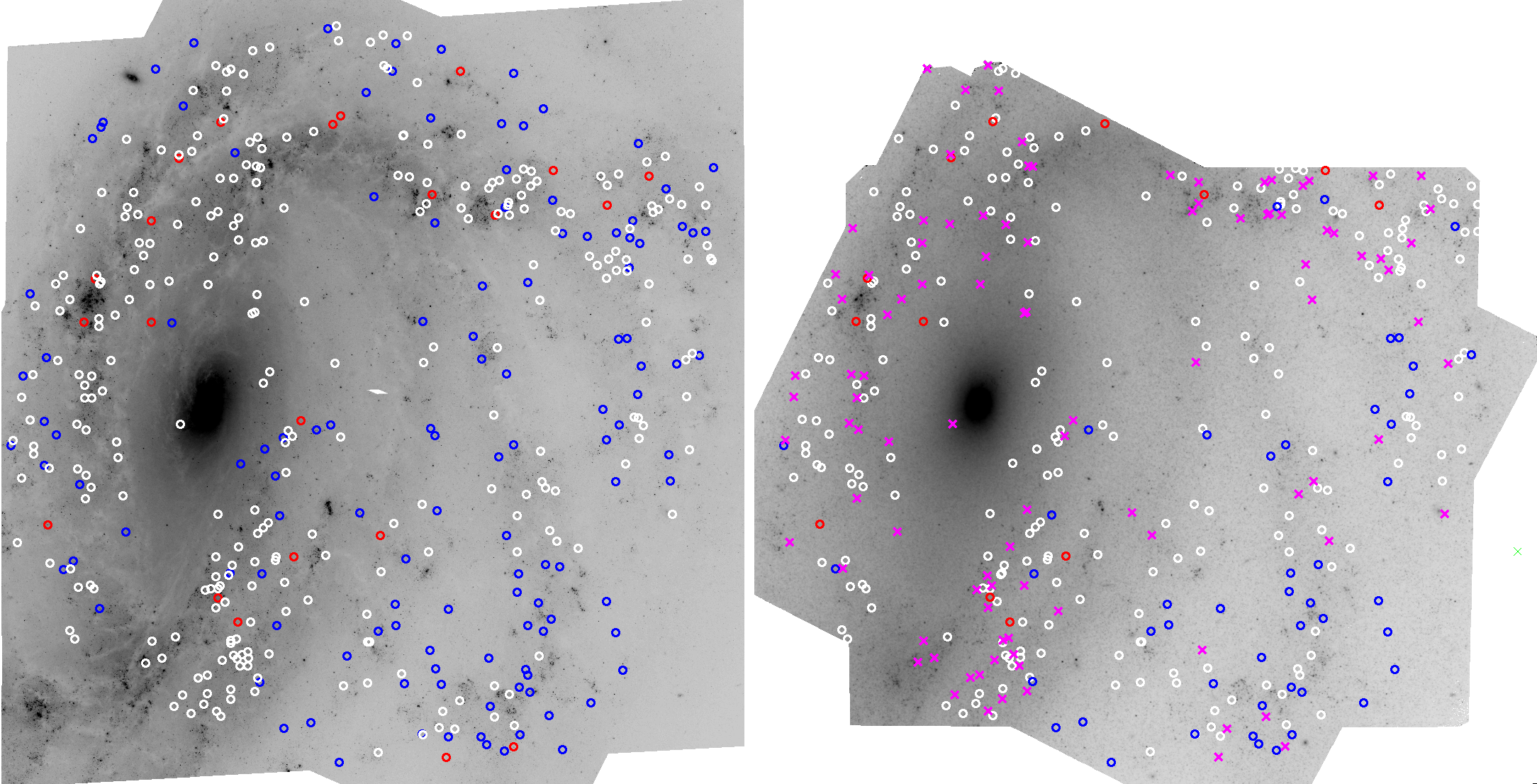}
\caption{Optical (left) and NIR (right) grayscale images of \ng showing the locations of Cepheids from various subsamples. Blue, white and red circles denote Cepheids used for distance determination, excluded variables, and rejected objects, respectively (matching the black, red and open symbols used in Fig.~\ref{fig_cs}). Magenta crosses in the right panel indicate optically-selected Cepheids that were not detected in the NIR frame.\label{fig_cephpos}}
\end{figure*}

\begin{figure*}
\includegraphics[width=\textwidth]{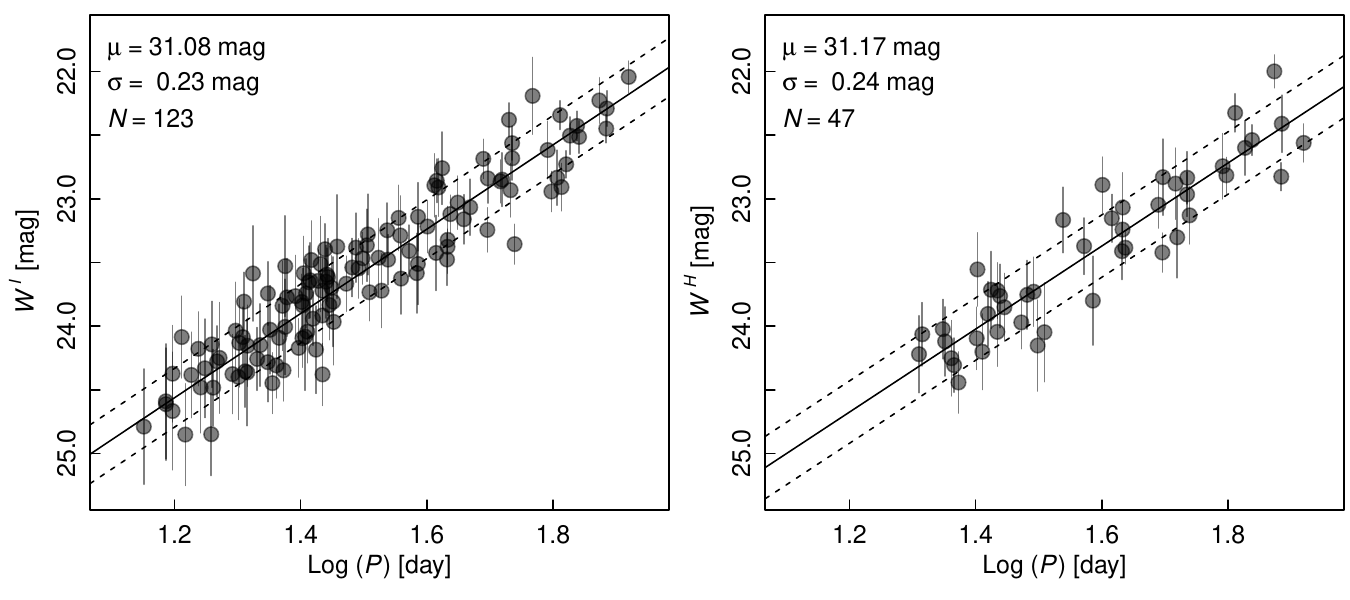}
\caption{Optical (left) and NIR (right) Wesenheit PLRs for the Cepheid subsamples used for distance determination. The solid lines indicate the best-fit PLRs, while the dashed lines indicate the 1$\sigma$ standard deviations.\label{fig_plr}}
\end{figure*}

These selection criteria are different from those adopted by Y20, where we applied cuts on period and \hsth surface brightness. Indeed, the crowding correction uncertainty is strongly correlated with period (or luminosity) and local surface brightness, as shown in Figure~\ref{fig_pss}. We had noted in \S4.3 of Y20 that a cut on $\sigma(\Delta\mathrm{\hsth})$ yielded results consistent with those obtained via a cut on local surface brightness. Fig.~\ref{fig_cephpos} shows the positions of all Cepheids, color-coded by subsample. We expect that a long-period or low-crowding Cepheid subsample will be minimally impacted by systematics.

\subsection{The distance to \ngs}\label{sec_eb}

We present the optical and NIR reddening-free Wesenheit PLRs for the distance-determination Cepheid subsamples in Figure~\ref{fig_plr}. The PLR fit results are summarized in Table~\ref{tab_plr}. The LMC-relative distance moduli are based on the Cepheid PLRs reported by \citet{Riess2019} and include a correction of $+0.021$~mag to our \hsth photometry to account for the difference in the adopted Vega-based zeropoint. The $W^I$ and $W^H$ PLRs yielded $\Delta\mu_I = 12.602 \pm 0.041$ and $\Delta\mu_H = 12.672 \pm 0.057$~mag, respectively. These two measurements are consistent within 1$\sigma$, and their combination yields $\Delta\mu = 12.626 \pm 0.033$~mag. We note the combined uncertainty of 0.033~mag can be slightly underestimated as $W^H$ and $W^I$ are partially correlated due to the color term in $W^H$. However, the \hsth term is independent of $W^I$ and contributes $2.6\times$ more than the color term. Thus, we conclude that taking the weighted mean of the $W^I$ and $W^H$ based distances is a reasonable approximation. \citet{1991PASP..103..933M}, \citet{2000A&A...356..849G} and \citet{2013ApJ...764...84I} suggested similar approaches that average distances from PLRs of different wavelengths. We adopt the LMC distance modulus of $\mu_\mathrm{LMC} = 18.477 \pm 0.026$~mag \citep{Pietrzynski2019} to arrive at a distance modulus for \ng of $\mu = 31.103 \pm 0.042$~mag, equivalent to a distance of $D = 16.62 \pm 0.32$~Mpc.

\begin{figure}
\includegraphics[width=0.49\textwidth]{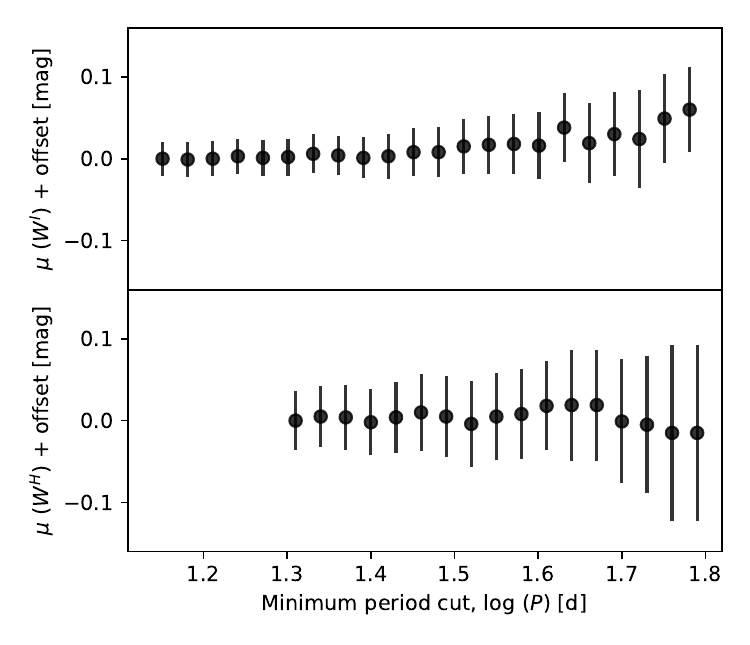}
\caption{Distance dependence on minimum period cut for the optical (upper) and NIR (lower) Cepheid subsamples.\label{fig_pcut}}
\end{figure}

We present our error budget in Table~\ref{tab_err} using the same format as for \ngo by Y20, with the addition of an error term from Cepheid period uncertainties. We derived the latter with Monte Carlo simulations. We randomly drew periods from Gaussian distributions whose standard deviations are determined with bootstrapping (see \S\ref{sec_cep}), then fit PLR to these simulations 1000 times. We found the resulting distance modulus errors due to period uncertainty are 0.006~mag and 0.010~mag for $W^I$ and $W^H$, respectively. Compared to our earlier work, we adopted a smaller and conservative magnitude calibration error of 0.01 mag for the optical PLR because we derived the \hsti and \hstv aperture corrections using field stars. The phase correction errors were also reduced as a result of the increased number of Cepheids in \ngs. The other entries are identical to \ngos. We estimated the uncertainty due to possible metallicity effects by using the [O/H] dispersion in 21 Cepheid hosts and the terms reported by \citet{Riess2016}. We refer interested readers to \S4.4 of Y20 for a detailed discussion of those systematics. We did not find a distance dependence on the minimum period cut for either the optical or NIR Cepheid subsamples, as shown in Figure~\ref{fig_pcut}.

We compared our distance determination to literature values, as compiled by the NASA/IPAC Extragalactic Database (NED). While the uncertainties in most of those previous determinations are too large to provide a statistically meaningful comparison against our result, \citet{2014ApJ...784L..11Y} derived a distance with comparable precision as ours. They used the AGN time-lag method and dust reverberation modeling to obtain a distance of $D = 17.6 \pm 0.6$~Mpc. Their result is consistent with ours within 1.4$\sigma$, though we note that their distance determination for \ngo is not consistent with ours, as reported by Y20.

\begin{deluxetable*}{cllrllcr}
\tabletypesize{\small}
\setlength{\tabcolsep}{10pt}
\tablecaption{Wesenheit PLRs used for distance determination\label{tab_plr}}
\tablewidth{0pt}
\tablehead{
\colhead{Index} & \multicolumn{2}{c}{Expression} & \multicolumn{1}{c}{Slope} & \multicolumn{2}{c}{Intercept}  & \multicolumn{1}{c}{Weighted} & \multicolumn{1}{c}{$N$} \\ \cline{5-6}
&& & & \multicolumn{1}{c}{LMC} & \multicolumn{1}{c}{N$\,$4051} & \multicolumn{1}{c}{scatter} & 
}
\startdata
$W^I$ & $\hsti - 1.3$   & $(\hstv-\hsti)$ & -3.31 & 15.933 & 28.535 & 0.23 & 123\\
$W^H$ & $\hsth - 0.386$ & $(\hstv-\hsti)$ & -3.26 & 15.915$^a$ & 28.587$^b$ & 0.24 & 47
\enddata
\tablecomments{$a$: Corrected for the \hsths-band count-rate nonlinearity; see Y20 for details. $b$: Adjusted by 0.021~mag for the difference in adopted Vega zeropoints between \citet{Riess2019} and this study.}
\end{deluxetable*}

\begin{deluxetable}{lll}
\tabletypesize{\normalsize}
\tablecaption{Error Budget\label{tab_err}}
\tablewidth{0pt}
\tablehead{
\colhead{Source} & \colhead{$W^I$} & \colhead{$W^H$} \\ 
& (mag) & (mag)
}
\startdata
PLR scatter & 0.021 & 0.035 \\
PLR slope & \nodata & 0.033 \\
Magnitude calibration & 0.01 & 0.02 \\
Metallicity & 0.03 & 0.02 \\
Phase correction & 0.009 & 0.004 \\
Reddening law & 0.01 & \nodata \\
Period error & 0.006 & 0.01\\
\hline
Subtotal & 0.041 & 0.057 \\
Subtotal combined & \multicolumn{2}{c}{0.033} \\
\hline
LMC distance & \multicolumn{2}{c}{0.026} \\
\hline
Total & 0.048 & 0.062 \\
Total combined & \multicolumn{2}{c}{0.042}
\enddata
\end{deluxetable}

\begin{figure}
\includegraphics[width=0.49\textwidth]{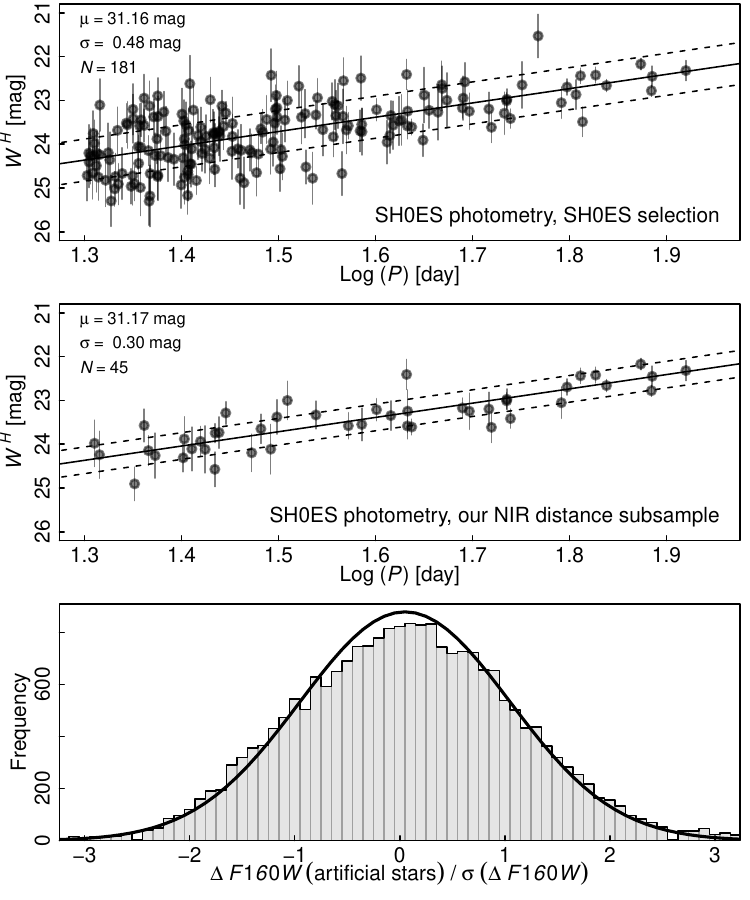}
\caption{{\it Top}: NIR Cepheid PLRs using SH0ES \hsth photometry and SH0ES selection criteria. The resultant distance modulus and weighted PLR scatter are shown in the top-left of the panel. {\it Middle}: Same as top but for our NIR calibration sample. {\it Bottom}: Error distribution of the SH0ES artificial star tests. For each artificial star, we subtracted the mean bias correction and divided by the standard deviation of all artificial stars associated with the same Cepheid.\label{fig_ss}}
\end{figure}

\subsection{Comparison to SH0ES procedure}

The SH0ES project \citep{Riess2016} measures Cepheid \hsth magnitudes following a different procedure, which gives us an opportunity to compare the two photometry approaches. We carried out SH0ES photometry on the \ng \hsth master image and derived NIR PLRs for a sample of Cepheids that were selected using the SH0ES criteria. Both methods use fixed-position PSF photometry in \hsths, with positions derived from the higher signal-to-noise ratio and better-sampled optical master frame. Likewise, they both use artificial star tests to correct for crowding. Unlike this study, where we fixed positions for all sources appearing in both \hstw and \hsth images, the SH0ES approach only fixes the positions for the Cepheids.

In addition, the SH0ES procedure includes two features to derive accurate photometry and Gaussian errors in magnitude space, as demonstrated in artificial star tests (see the bottom panel of Figure~\ref{fig_ss}): (1) a large ``protection'' radius of $0.75\times$ FWHM to avoid spurious de-blends; (2) the selection of artificial stars that match the apparent source displacement of the Cepheid relative to the predicted optical position \citep[a secondary consequence of blending,][]{2009ApJ...699..539R}. As a result, the SH0ES approach yielded unbiased photometry for a broader range of local crowding levels and thus resulted in a larger sample of Cepheids for distance determination. As shown in the top panel of Figure~\ref{fig_ss}, the SH0ES criteria yielded 182 Cepheids and a distance modulus of 31.16~mag using the same aforementioned PLR slope. We also determined a distance modulus using the SH0ES-based photometry for 45 Cepheids that are in common with our NIR subsample, which yielded 31.17~mag (middle panel of Figure~\ref{fig_ss}). Both are in excellent agreement (to within $0.01$~mag) with our results. We expect a statistical uncertainty between methods mostly due to the finite number of unique samples of artificial stars, $\sim$0.006~mag, and a systematic uncertainty arising from the choice of different PSFs, sky background regions, fitting techniques, etc., $\sim$0.02~mag.

This comparison is only a cross-check of the \hsth photometry and the NIR subsample selection between the SH0ES approach and the methods adopted in this work. We refer interested readers to \citet{2021arXiv210212489J} for a comprehensive and fully independent comparison of all steps in the SH0ES methodology.

\section{Discussion}

Our Cepheid-based distance to \ng is close to the high end of the literature values, which range from $\sim$9 to 18 Mpc and are mostly based on the \citet{Tully1977} relationship.  The bright central AGN in \ng is generally not accounted for in Tully--Fisher studies, thus biasing the studies that use blue imaging towards smaller distances (Robinson et al., submitted).  Additionally, \ng may have increased star formation, when compared to comparable quiescent galaxies, as has been found for many active galaxies in the local universe \citep{2010ApJ...720..368X}. An elevated star formation rate may lead to an elevated surface brightness \citep[e.g.,][]{2010ApJ...717..803G,2020FrASS...7...21M}, which would also serve to bias Tully--Fisher distances towards smaller values.  Finally, \ng has a lower inclination than is generally recommended for Tully--Fisher studies and thus increases the uncertainty in the resultant distance \citep{2020ApJ...902..145K}.  We discuss the implications of the Cepheid-based distance in \S5.1 \& \S5.2, and the \ng galactic motion in \S5.3.

\subsection{Eddington ratio in \ngs}

\begin{figure}
\includegraphics[width=0.49\textwidth]{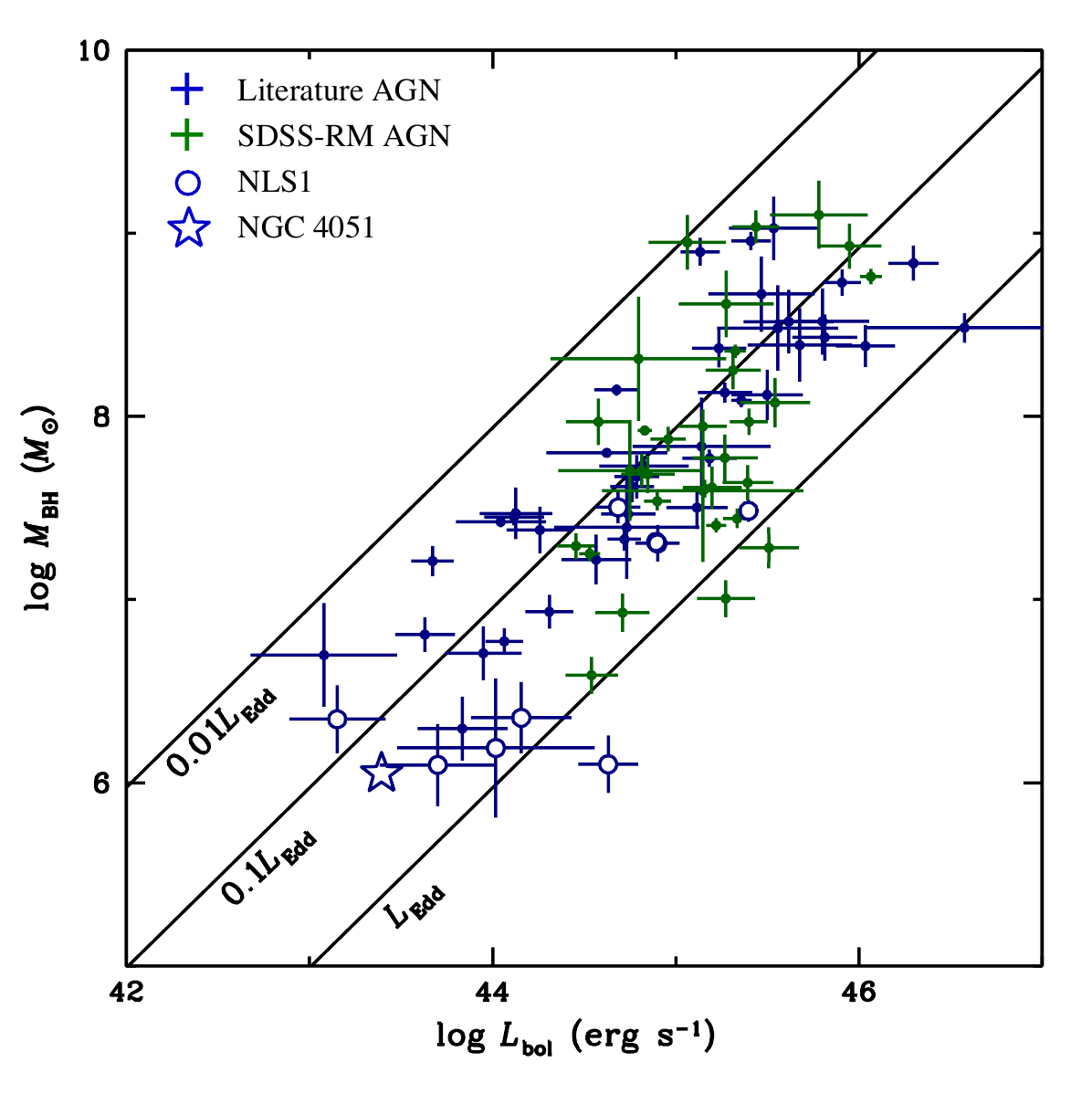}
\caption{Black hole mass versus bolometric luminosity for AGNs with reverberation-based masses using the \hb\ emission line. Shown in blue are sources gathered from the literature by \citet[their Table A1]{DB2020}; shown in green are SDSS-RM AGNs with \hb-based reverberation masses \citep{Grier2017} as compiled by \citet[their Table A2]{DB2020}. NLS1s are shown as open circles. The updated measurement for \ng is represented by the open star in the lower left of the diagram.
\label{fig_ML}}
\end{figure}

\cite{Peterson2004} and \cite{Collin2006} recognized that there might be a problem with the redshift-based distance to \ngs, and it was these concerns that led us to this investigation. \citet[their Figure 16]{Peterson2004} and \citet[their Figure 6]{Collin2006} showed that the Eddington ratio in \ng was surprisingly low for an NLS1 galaxy, one of only two such objects with $\dot{m} < 0.1$ \citep[the other is MCG-06-30-15;][]{Bentz2016a}. Now, at the corrected distance of 16.6\,Mpc, the Eddington ratio is $\dot{m} \approx 0.2$, a value more typical of NLS1 galaxies. In Figure~\ref{fig_ML}, we update the mass--luminosity diagram of \citet{Peterson2004}, where we see that the improved and increased distance places \ng in the same locus as other NLS1s, with $\dot{m} \gtrsim 0.1$, instead of the seemingly-odd value implied by previous distance estimates.

\subsection{Revisiting the energetics of the nucleus}

It seems that the most commonly assumed distance for \ng
is the redshift-based distance of 9.3\,Mpc, which is considerably
smaller than the value reported here. This has a significant effect on linear distances within the galaxy, which now become a factor of 1.8 larger, and luminosity measures, which increase by a factor of 3.2.

\cite{Barbosa2009}, for example, estimated the mass outflow rate and power along with three other Seyfert galaxies. Repeating their outflow calculation and increasing the diameter of the eastern outflow from 45\,pc to 80\,pc results in an outflow rate $\dot{M} = 2.2\,n_{100}f\, \msun\,{\rm yr}^{-1}$,
where $n_{100}$ is the electron density in units of $100\,{\rm cm}^{-3}$ and
$f$ is the volume filling factor, for which \cite{Barbosa2009} suggest a value of $10^{-3}$. We similarly correct the diameter of the western outflow from 44.5\,pc to 79.2\,pc, thus increasing the outflow rate to $1.4\,n_{100}f\,\msun\,{\rm yr}^{-1}$. The total outflow rate is the sum of these, $\dot{M}= 3.6\,n_{100}f\,\msun\,{\rm yr}^{-1}$. Similarly, recalculation of the kinetic luminosity gives
\begin{equation}
P=\frac{1}{2}\dot{M}V^2 = 2.6\times10^{40}n_{\rm 100}f\,\ergsec,
\end{equation}
where $V$ is the outflow velocity. 

\cite{Barbosa2009} compare the mass outflow rate to the mass accretion rate required to sustain the luminosity, but unfortunately using a bolometric luminosity that is much too low. The luminosity they use traces back to a {\em Chandra} X-ray observation obtained in 2000 April, when \ng was
in an extremely faint X-ray state \citep{Collinge2001}; \ng has some history of the X-rays nearly turning off for long periods of time \citep{Uttley1999,Peterson2000}. We instead use the bolometric luminosity from \cite{EDB2020}, which adjusted to our distance of 16.6\,Mpc becomes $\log L_{\rm bol} = 43.38$. It is worth noting at this point that the kinetic luminosity of the outflow is some six orders of magnitude less than the radiative luminosity for $f\sim10^{-3}$.

The corrected mass accretion rate is thus
\begin{equation}
\dot{M}_{\rm acc} = \frac{L_{\rm bol}}{\eta\,c^2} = 
0.0042\,\msun\,{\rm yr}^{-1},
\end{equation}
where we have taken the efficiency to have the nominal value $\eta = 0.1.$ \cite{Barbosa2009} argue that if the total outflow rate exceeds the accretion rate, then the outflow must be comprised of accelerated circumnuclear material. In our revised calculation, this condition is met if $f \gtrsim 10^{-3}n_{100}^{-1}$.

On the other hand, the analysis of high-velocity outflows in \ng by \cite{Pounds2014} does not need reassessment as a distance of 15.2\,Mpc, based on a Tully--Fisher measurement, was assumed \citep{Pounds2011a,Pounds2011b,Pounds2013}. Adjustment to a distance of 16.6\,Mpc results in little change.

\subsection{Galactic motion}

\ng is a member of what is often referred to as the Ursa Major Cluster \citep{1996AJ....112.2471T}. {\it CosmicFlows-3} \citep{2016AJ....152...50T} reports an overall distance to Ursa Major of $17.2\pm1.0$~Mpc, which agrees with our Cepheid distance to \ng within the quoted uncertainty. Over the years, it has become clear that the Ursa Major complex is likely composed of several distinct bound groups \citep{2013MNRAS.429.2264K, 2017ApJ...843...16K}. Both \citet{2013MNRAS.429.2264K} and \citet{2017ApJ...843...16K} place \ng into a group with the lenticular galaxy NGC\,4111 as the brightest member. \citet{2017ApJ...843...16K} list 14 members in the group, and have redshift-independent distances for five of these galaxies. With this information, they derive a weighted average {\it group} distance of 14.36 $\pm$ 1.01 Mpc, which is broadly consistent with our derived Cepheid distance to \ng when taking into account the size of the group itself.

The observed velocity of \ng in the reference frame of the Local Group is $v_\mathrm{LG} = 740 \pm 3$~km~s$^{-1}$ \citep{1996AJ....111..794K}. Adopting a value of the Hubble constant of $74.0\pm 1.4$~km~s$^{-1}$~Mpc$^{-1}$ \citep{Riess2019} allows us to calculate a peculiar velocity of $v_\mathrm{pec} = -490 \pm 34$~km~s$^{-1}$. While seemingly high, we note that the vast majority of galaxies in the Ursa Majoris complex exhibit a negative peculiar velocity, with an average peculiar velocity with respect to the Local Group of $v_\mathrm{pec} = -337$ km~s$^{-1}$ \citep{2013MNRAS.429.2264K}. Similarly, \citet{2014MNRAS.445..630P} identified 166 Ursa Major members and found that they lie in a redshift range of 620--1208~km~s$^{-1}$. Considering that all these galaxies are roughly equidistant, the peculiar velocity of \ng lies well within the inferred peculiar velocity range of other Ursa Major galaxies. When placed in this context, our derived distance to \ng is consistent with expectations.

\section{Summary}

We analyzed \hst time-series observations for the Seyfert 1 galaxy \ng and identified 419 Cepheid candidates in this system. Using subsamples of Cepheids that were precisely measured at optical and NIR wavelengths, we derived a distance of $D = 16.6 \pm 0.3$~Mpc, equivalent to a distance modulus of $\mu=31.10\pm0.04$~mag. We calculated the Eddington ratio of the AGN residing in this galaxy to be $\dot{m} \approx 0.2$, thus confirming \ng as a typical NLS1 in terms of the AGN mass--luminosity ratio. We obtained a peculiar velocity of $-490 \pm 34$~km~s$^{-1}$ for \ng and compared it to those of the other Ursa Major member galaxies. We revisited the energetics of the \ng nucleus based on \citet{Barbosa2009} and derived an outflow rate of $\sim 3.6\times 10^{-3}\,\msun\,{\rm yr}^{-1}$ and an outflow kinetic luminosity of $\sim 2.6\times 10^{37}\,{\rm erg}\,{\rm s}^{-1}$ if $n_{100} f\sim10^{-3}$ was assumed, as well as a mass accretion rate of $\sim 4.2\times 10^{-3}\,\msun\,{\rm yr}^{-1}$.

\acknowledgments

We thank K.Z.~Stanek for helpful advice early in this project, and the anonymous referee for providing valuable comments. MMF and BMP thank Tod Lauer for an enlightening conversation. Support for {\em HST} program GO-13765 was provided by NASA through a grant from the Space Telescope Science Institute, which is operated by the Association of Universities for Research in Astronomy, Inc., under NASA contract NAS5-26555. MCB gratefully acknowledges support from the NSF through grant AST-2009230. MV gratefully acknowledges support from the Independent Research Fund Denmark via grant number DFF 8021-00130. This research has made use of the NASA/IPAC Extragalactic Database (NED), which is funded by the National Aeronautics and Space Administration and operated by the California Institute of Technology.

\appendix

\section{Supplementary Material}

We include light curves (in electronic form) for 2804 objects that passed our initial cuts (prior to visual selection). Light curves are shown in the same form as one of the panels of Figure~\ref{fig_lc}. The file names are in the format of {\tt <id>.pdf} where {\tt <id>} is our internal tracking identifier. The figures are compressed and grouped into two tarballs, {\tt selected.tar.gz} and {\tt rejected.tar.gz}, based on our visual inspection results.

\bibliographystyle{aasjournal}
\bibliography{ref}
\end{document}